\documentclass{article}
\PassOptionsToPackage{numbers,sort&compress}{natbib}
\usepackage[preprint,nonanonymous]{neurips_2026}
\makeatletter
\renewcommand{\@notice}{}
\makeatother
\usepackage[utf8]{inputenc}
\usepackage[T1]{fontenc}
\usepackage{hyperref}
\usepackage{url}
\usepackage{booktabs}
\usepackage{amsmath}
\usepackage{amssymb}
\usepackage{graphicx}
\usepackage{subcaption}
\usepackage{multirow}

\title{Auditing Discriminatory Patterns in Mortgage Lending Through Association Rules and Fair Binning}

\author{
  Archit Rathod \\
  University of Illinois Chicago \\
  \texttt{arath21@uic.edu} \\
  \And
  Dhwani Chande \\
  University of Illinois Chicago \\
  \texttt{dchan35@uic.edu} \\
  \And
  Het Nagda \\
  University of Illinois Chicago \\
  \texttt{hnagd@uic.edu} \\
}

\begin{document}
\maketitle

\begin{abstract}
Mortgage lending in the United States exhibits persistent racial and gender disparities.
We investigate whether standard data preprocessing steps, specifically attribute binning, amplify these disparities in downstream pattern mining.
Using 103,481 cleaned mortgage applications from the HMDA 2023 dataset (Chicago metropolitan area), we build a three-stage pipeline: (1) a PySpark data cleaning and binning pipeline that implements both standard equal-frequency binning and the $\varepsilon$-biased fair binning algorithm from Asudeh et al.~\cite{asudeh2025}, (2) FP-Growth association rule mining that compares denial patterns under both binning regimes, and (3) K-Means clustering with a per-cluster disparate impact audit.
Our standard binning shows 9.63\% racial bias in income discretization, consistent with the 8--10\% reported in prior work.
Fair binning with seven race groups is infeasible at $\varepsilon{=}0.03$ and only succeeds at $\varepsilon{=}0.08$ with a Price of Fairness of 29.4\%.
FP-Growth reveals that high debt-to-income ratio is the dominant denial predictor (67.2\% confidence, 2.81 lift), while racial bias does not appear as explicit high-support rules.
However, K-Means clustering followed by a disparate impact audit flags 10 out of 45 cluster-group pairs, showing that Black applicants face significantly higher denial rates than White applicants even among financially similar groups.
\end{abstract}

\section{Introduction}

The Home Mortgage Disclosure Act (HMDA) requires nearly every mortgage lender in the U.S. to report detailed information about each loan application, including the applicant's race, sex, income, and the lending decision.
The Consumer Financial Protection Bureau (CFPB) publishes this data annually and uses it for enforcement actions against discriminatory lenders~\cite{cfpb2023hmda}.
The dataset contains over 10 million records per year with 99 fields per application, making it one of the richest public sources for studying lending discrimination.

Prior work has focused on building fair classifiers or auditing model-level bias~\cite{barocas2016big, mehrabi2021survey, angwin2016machine}.
However, upstream data management operations can also introduce bias~\cite{stoyanovich2020responsible}.
Recent work on fairness-aware data systems, including data structures such as FairHash, reinforces that fairness concerns can arise throughout the data pipeline and not only at the model layer~\cite{shahbazi2024fairhash}.
Asudeh et al.~\cite{asudeh2025} recently showed that standard equal-frequency binning of numeric attributes like income can produce bins with significantly skewed demographic compositions, amplifying existing disparities.
They proposed the $\varepsilon$-biased binning algorithm that bounds demographic deviation within each bin to at most $\varepsilon$.

Our project connects these ideas by asking: \textit{does fairness-aware binning change the discriminatory patterns mined from lending data?}

We approach this question through three complementary methods.
First, we implement both standard and $\varepsilon$-biased binning on HMDA data and measure the resulting bias.
Second, we use FP-Growth~\cite{han2000mining} to mine association rules predicting loan denial under both binning regimes and compare the rule sets.
Third, we cluster applicants by financial profile using K-Means and compute the Disparate Impact Ratio (DIR) within each cluster to identify discrimination that cannot be explained by legitimate risk factors.

\section{Problem Formalization}

\subsection{Data Model}

Let $\mathcal{D} = \{t_i\}_{i=1}^{n}$ be a set of $n$ mortgage applications, each described by attributes $X = \{x_1, \ldots, x_d\}$.
Each tuple $t$ belongs to a demographic group $g(t) \in G = \{g_1, \ldots, g_\ell\}$.
In our setting, $G$ represents racial categories (White, Black, Asian, etc.), and the outcome variable $y \in \{0, 1\}$ indicates whether the application was denied ($y{=}1$) or originated ($y{=}0$).

\subsection{Binning and Bias}

A $k$-binning of a numeric attribute $x$ partitions its values into $k$ ordered buckets $\mathcal{B} = \{B_1, \ldots, B_k\}$ by specifying $k{-}1$ boundaries.
Equal-frequency binning, a standard discretization technique in data mining, ensures each bucket has approximately $n/k$ tuples~\cite{kotsiantis2006discretization}.

Following Asudeh et al.~\cite{asudeh2025}, we define the bias of a bucket $B_j$ with respect to group $g_l$ as:
\begin{equation}
\beta(B_j, g_l) = \left|\frac{|G_l \cap B_j|}{|B_j|} - \frac{|G_l|}{n}\right|
\label{eq:bin_bias}
\end{equation}
This measures how much the group's proportion within a bin deviates from its overall proportion in the dataset.
The overall binning bias is the worst case across all bins and groups:
\begin{equation}
\beta(\mathcal{B}) = \max_{l \in [\ell],\; j \in [k]} \beta(B_j, g_l)
\label{eq:overall_bias}
\end{equation}

A binning is $\varepsilon$-biased if $\beta(\mathcal{B}) \leq \varepsilon$.
The goal is to find the $\varepsilon$-biased $k$-binning that minimizes the difference between the largest and smallest bin sizes (the Price of Fairness).

\subsection{Association Rule Mining}

Given a set of transactions (discretized application records), frequent itemset mining identifies combinations of items that appear together in at least a fraction (support) of all transactions.
An association rule $A \Rightarrow C$ has confidence $P(C | A)$ and lift $P(C | A) / P(C)$.
We look for rules of the form $\{$protected attribute, financial features$\} \Rightarrow \{$denied$\}$ to surface discriminatory patterns.

\subsection{Disparate Impact Ratio}

The four-fifths rule from U.S. fair lending law states that a lending practice has disparate impact if the acceptance rate of a protected group is less than 80\% of the acceptance rate of the reference group~\cite{eeoc1978uniform}.
We formalize this as:
\begin{equation}
\text{DIR}(g, g_{\text{ref}}) = \frac{1 - \text{denial\_rate}(g)}{1 - \text{denial\_rate}(g_{\text{ref}})}
\label{eq:dir}
\end{equation}
A DIR below 0.8 indicates disparate impact.
By computing DIR within clusters of financially similar applicants, we isolate discrimination from legitimate risk factors.

\section{Dataset and Preprocessing}

\subsection{Data Acquisition}

We used the HMDA 2023 Snapshot National Loan-Level Dataset published by the CFPB at \url{https://ffiec.cfpb.gov/data-publication}.
The Snapshot is a frozen copy of all national lending data (freeze date: May 1, 2024), which ensures reproducibility.
The raw file contains approximately 10 million rows and 99 columns as a 4 GB CSV.
We loaded the file using PySpark on Google Colab (T4 GPU, 10 GB driver memory) and filtered to the Chicago-Naperville-Elgin Metropolitan Division (code 16984), yielding 204,100 records.

\subsection{Data Cleaning}

We applied five sequential cleaning steps, shown in Table~\ref{tab:cleaning}.

\begin{table}[htbp]
\centering
\caption{Data cleaning pipeline. Each step filters the dataset further.}
\label{tab:cleaning}
\begin{tabular}{clr}
\toprule
\textbf{Step} & \textbf{Operation} & \textbf{Rows After} \\
\midrule
0 & Filter to Chicago MSA 16984 & 204,100 \\
1 & Keep only originated (code 1) and denied (code 3) & 134,894 \\
2 & Remove unavailable race, sex, ethnicity & 106,754 \\
3 & Cast financial fields; drop null/invalid income & 104,477 \\
4 & Clip income outliers (0.5th--99.5th percentile) & 103,481 \\
\bottomrule
\end{tabular}
\end{table}

HMDA stores financial fields as strings with special values ``NA'' and ``Exempt.'' We explicitly mapped these to null values using PySpark \texttt{F.when()} chains before casting to double.
The debt-to-income ratio (DTI) is reported as ranges (e.g., ``20\%-<30\%'') and individual integers (e.g., ``36''), which we mapped to midpoint numeric values for clustering.
Income outliers were clipped at the 0.5th and 99.5th percentiles (\$16K--\$1,025K) to remove data entry errors while preserving 99\% of records.

\paragraph{Schema-specific issues.}
The HMDA 2023 schema uses \texttt{derived\_msa\_md} rather than \texttt{msa\_md}, and Chicago's Metropolitan Division code is \texttt{16984}, not the broader MSA code \texttt{16980}.
The field \texttt{property\_type} does not exist in the schema; the correct field is \texttt{derived\_dwelling\_category}.
Spark cannot read \texttt{.zip} files natively, so we extracted the CSV using Python's \texttt{zipfile} module before loading.

\subsection{Cleaned Dataset Summary}

The final dataset contains 103,481 mortgage applications with 20 columns.
The overall denial rate is 23.9\%.
Table~\ref{tab:demographics} shows denial rates across demographic groups.

\begin{table}[htbp]
\centering
\caption{Denial rates by race and income quartile in the cleaned Chicago dataset.}
\label{tab:demographics}
\begin{tabular}{lrrr}
\toprule
\textbf{Race} & \textbf{Count} & \textbf{\% of Total} & \textbf{Denial Rate} \\
\midrule
White & 73,799 & 71.3\% & 20.9\% \\
Black or African American & 16,171 & 15.6\% & 38.6\% \\
Asian & 10,149 & 9.8\% & 21.6\% \\
Joint & 2,219 & 2.1\% & 15.1\% \\
American Indian or Alaska Native & 691 & 0.7\% & 40.8\% \\
2 or more minority races & 281 & 0.3\% & 49.1\% \\
Native Hawaiian or Other Pacific Islander & 171 & 0.2\% & 45.6\% \\
\bottomrule
\end{tabular}

\vspace{6pt}

\begin{tabular}{lrrr}
\toprule
\textbf{Income Quartile} & \textbf{Count} & \textbf{Avg Income (\$K)} & \textbf{Denial Rate} \\
\midrule
Q1 Low & 26,305 & 53 & 35.3\% \\
Q2 Mid-Low & 25,722 & 87 & 24.0\% \\
Q3 Mid-High & 25,627 & 129 & 21.0\% \\
Q4 High & 25,827 & 278 & 15.1\% \\
\bottomrule
\end{tabular}
\end{table}

Black applicants have an average income of \$99K compared to \$139K for White applicants, a 29\% gap.
This income difference means that when income is binned without considering demographics, Black applicants are disproportionately concentrated in lower bins.

\section{Methods}

\subsection{Standard Equal-Frequency Binning}

We binned \texttt{income} and \texttt{loan\_amount} into $k{=}5$ equal-frequency buckets using \texttt{pd.qcut()}, producing bins with approximately 20,700 records each.
We measured binning bias using Equations~\ref{eq:bin_bias} and~\ref{eq:overall_bias}.

\subsection{\texorpdfstring{$\varepsilon$-Biased Fair Binning}{Epsilon-Biased Fair Binning}}

We implemented the Divide-and-Conquer (D\&C) algorithm (Algorithm 5 from Asudeh et al.~\cite{asudeh2025}) for $\varepsilon$-biased binning.
The algorithm works by:

\begin{enumerate}
\item Sorting all records by the attribute value and building a prefix count array for each demographic group, enabling $O(1)$ bias checks for any segment.
\item Starting at the midpoint of the equal-frequency partition and checking whether splitting there produces two $\varepsilon$-biased super-segments.
\item If not, searching outward ($\pm 1, \pm 2, \ldots$) until a valid boundary is found.
\item Recursing on both halves until all $k$ bins are formed.
\end{enumerate}

The algorithm runs in $O(n \log k)$ time and finds a valid solution whenever one exists (Theorem 4 in~\cite{asudeh2025}).

We also computed the Price of Fairness (PoF), which measures how much the fair bin sizes deviate from equal-frequency:
\begin{equation}
\text{PoF}(\mathcal{B}) = \frac{1}{k} \sum_{j=1}^{k} \left|1 - \frac{k \cdot |B_j|}{n}\right|
\label{eq:pof}
\end{equation}

\subsection{FP-Growth Association Rule Mining}

Each application was converted into a transaction of 11 categorical items: outcome (originated/denied), race, sex, ethnicity, age bracket, income bin, loan amount bin, DTI category, loan type, loan purpose, and occupancy type.
Column values were prefixed with the feature name to create unique items (e.g., \texttt{race=White}, \texttt{dti\_bin=DTI\_High}).

We used the FP-Growth implementation from \texttt{mlxtend}~\cite{raschka2018mlxtend} with the following parameters: minimum support of 0.10, maximum itemset length of 4, minimum confidence of 0.50, and minimum lift of 1.0.
FP-Growth was run on both the standard-binned and fair-binned transactional datasets.

\subsection{K-Means Clustering and Disparate Impact Audit}

We clustered applicants using PySpark MLlib's K-Means on four financial features: \texttt{income}, \texttt{loan\_amount}, \texttt{dti\_numeric}, and \texttt{combined\_loan\_to\_value\_ratio}.
Interest rate was deliberately excluded because HMDA records this field only for originated loans; including it would have filtered out every denied application and made the disparate-impact audit impossible.
Features were z-score standardized using \texttt{StandardScaler} (with mean centering and unit variance) so that all features contribute equally to the Euclidean distance computation.

Rows with null values in any clustering feature were dropped (3,957 rows, 3.8\%), leaving 99,524 records for clustering.
We verified that the denial rate in this modeling frame (23.5\%) closely matches the full sample (23.9\%), confirming that the dropped rows do not introduce outcome bias.

We selected the optimal $k$ by testing $k \in \{2, \ldots, 8\}$ and evaluating both Within-Cluster Sum of Squares (WCSS) and silhouette score.
We enforced a floor of $k \geq 4$ because 2- or 3-cluster solutions are too coarse for a meaningful DIR audit.

After clustering, we computed the DIR for each cluster-group pair using the EEOC selection-rate formulation defined in Equation~\ref{eq:dir}, where DIR is the protected group's acceptance rate divided by the reference group's acceptance rate.
A DIR below 0.80 indicates disparate impact under the four-fifths rule.
We audited three demographic attributes: race (reference: White), ethnicity (reference: Not Hispanic or Latino), and sex (reference: Male).
Groups with fewer than 30 members in a cluster were excluded to avoid unreliable ratios.

\section{Results}

\subsection{Binning Bias}

Table~\ref{tab:binning_bias} compares the bias of standard and $\varepsilon$-biased binning.

\begin{table}[htbp]
\centering
\caption{Binning bias comparison. Bias is measured as the maximum demographic deviation across all bins and groups (Equation~\ref{eq:overall_bias}).}
\label{tab:binning_bias}
\begin{tabular}{llrrrr}
\toprule
\textbf{Attribute} & \textbf{Method} & \textbf{Race Bias} & \textbf{Sex Bias} & \textbf{$\varepsilon$ Used} & \textbf{PoF} \\
\midrule
Income & Standard & 9.63\% & 25.14\% & -- & 1.48\% \\
Income & $\varepsilon$-biased & 8.00\% & 21.08\% & 0.08 & 29.43\% \\
Loan Amount & Standard & 7.65\% & -- & -- & 4.28\% \\
Loan Amount & $\varepsilon$-biased & 6.00\% & -- & 0.06 & 23.54\% \\
\bottomrule
\end{tabular}
\end{table}

The standard income binning produces 9.63\% racial bias, consistent with the 8--10\% reported by Asudeh et al. on credit scoring data~\cite{asudeh2025}.
With all seven race groups, the D\&C algorithm could not find a feasible solution at $\varepsilon{=}0.03$.
It also failed at $\varepsilon{=}0.04$ and $0.05$.
For income, a feasible binning was found only at $\varepsilon{=}0.08$, and for loan amount, at $\varepsilon{=}0.06$.
The resulting Price of Fairness is high (29.4\% and 23.5\%), meaning the bin sizes deviate substantially from equal-frequency.

This outcome is consistent with Figures 6b and 6f of the paper~\cite{asudeh2025}, which show that increasing the number of groups sharply reduces the set of valid boundary candidates and increases infeasibility.

\begin{figure}[htbp]
\centering
\includegraphics[width=\textwidth]{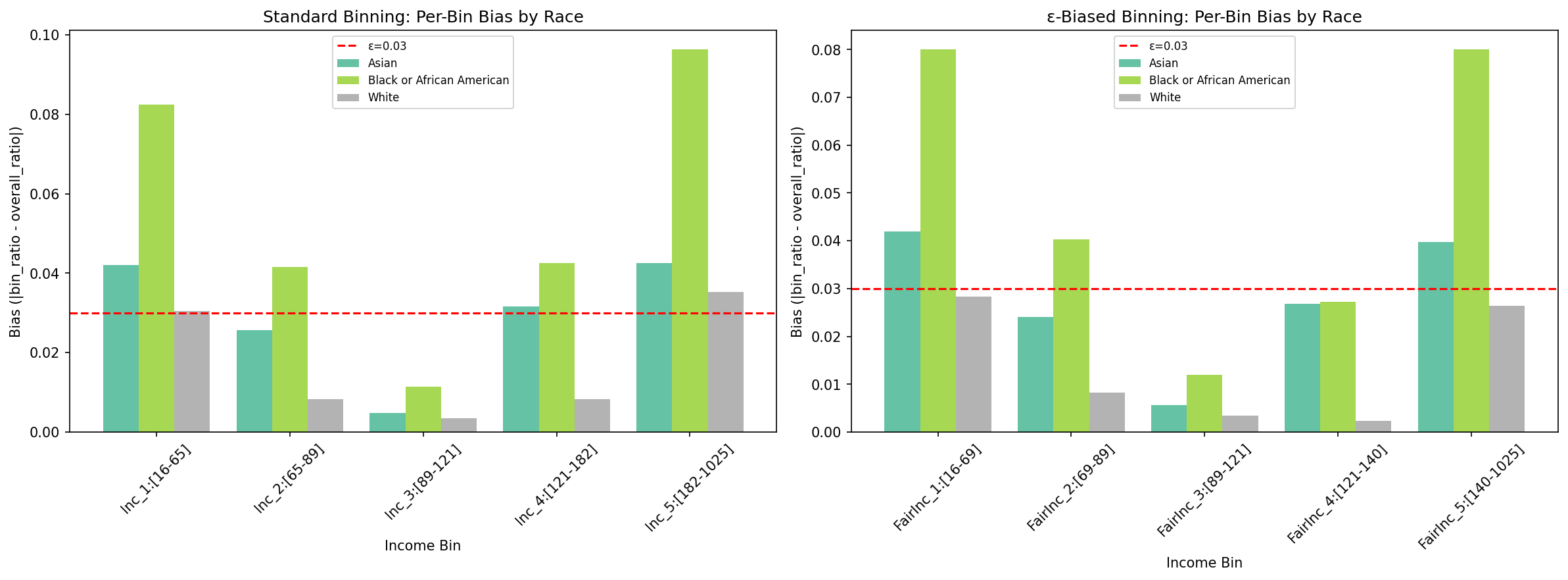}
\caption{Per-bin race bias under standard (left) vs. $\varepsilon$-biased (right) income binning. The red dashed line marks the target $\varepsilon{=}0.03$.}
\label{fig:binning_comparison}
\end{figure}

\subsection{FP-Growth Results}

Table~\ref{tab:fpgrowth} summarizes the FP-Growth results under both binning regimes.

\begin{table}[htbp]
\centering
\caption{FP-Growth comparison: standard vs. fair binning.}
\label{tab:fpgrowth}
\begin{tabular}{lrr}
\toprule
\textbf{Metric} & \textbf{Standard} & \textbf{Fair} \\
\midrule
Frequent itemsets & 819 & 854 \\
Association rules (conf $\geq$ 0.50, lift $\geq$ 1.0) & 2,214 & 2,454 \\
Denial-consequent rules & 3 & 3 \\
Rules with demographic antecedents & 0 & 0 \\
Average confidence & 0.754 & 0.738 \\
Average lift & 1.140 & 1.208 \\
\bottomrule
\end{tabular}
\end{table}

The three denial rules are identical under both binning regimes (Table~\ref{tab:denial_rules}).
All are driven by high debt-to-income ratio, not by income or loan amount bins.
Since the fair binning only changed income and loan amount bin boundaries, the denial rules are unaffected.

\begin{table}[htbp]
\centering
\caption{Top denial rules (identical under standard and fair binning).}
\label{tab:denial_rules}
\begin{tabular}{lcccc}
\toprule
\textbf{Rule} & \textbf{Support} & \textbf{Confidence} & \textbf{Lift} \\
\midrule
$\{\text{DTI\_High}\} \Rightarrow \{\text{denied}\}$ & 0.107 & 0.672 & 2.81 \\
$\{\text{DTI\_High, PrimaryRes}\} \Rightarrow \{\text{denied}\}$ & 0.102 & 0.663 & 2.78 \\
$\{\text{DTI\_High}\} \Rightarrow \{\text{PrimaryRes, denied}\}$ & 0.102 & 0.637 & 2.82 \\
\bottomrule
\end{tabular}
\end{table}

No rules with race, sex, or ethnicity in the antecedent reached the 10\% support threshold.
This does not indicate the absence of racial bias.
Rather, it means that racial bias operates indirectly through correlated financial features.
Black applicants' average income (\$99K) is 29\% lower than White applicants' (\$139K), so they are disproportionately affected by income-related denial patterns.
The 10\% support threshold is too high for minority groups to appear: Black applicants comprise 15.6\% of the data, and denied Black applicants represent roughly 6\% (15.6\% $\times$ 38.6\%).

\begin{figure}[htbp]
\centering
\includegraphics[width=\textwidth]{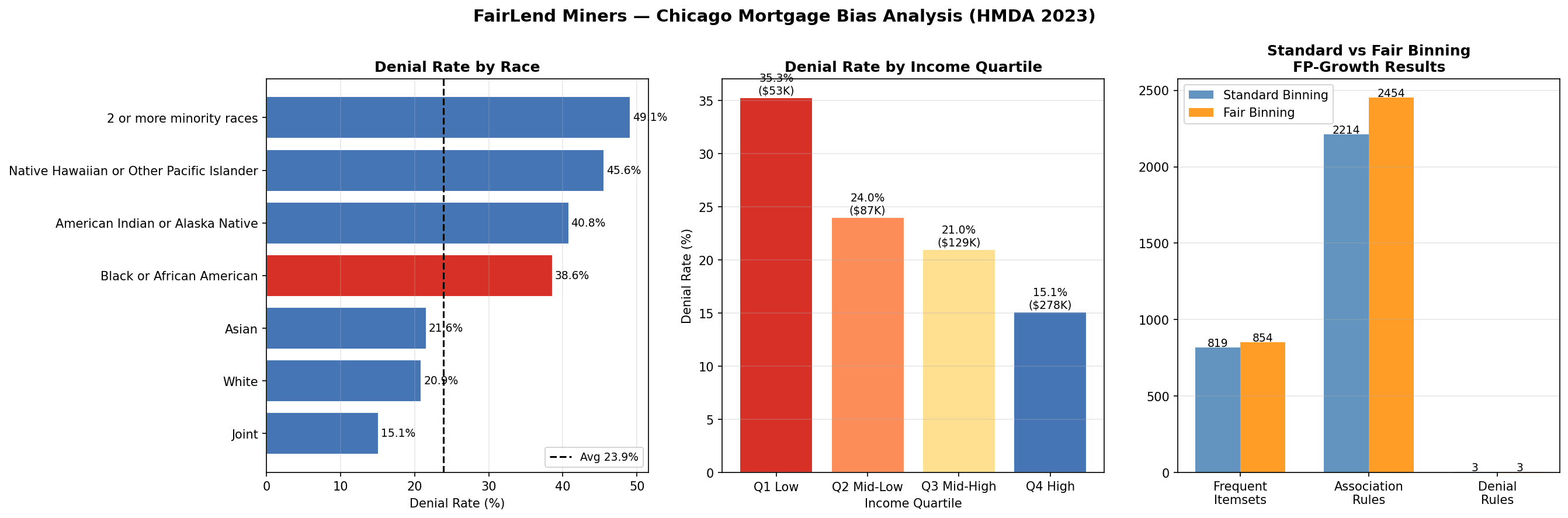}
\caption{Left: denial rate by race. Center: denial rate by income quartile. Right: standard vs. fair binning FP-Growth comparison.}
\label{fig:memberB}
\end{figure}

\subsection{Clustering Results}

Table~\ref{tab:k_selection} shows the WCSS and silhouette scores for each candidate $k$.
We selected $k{=}5$ (silhouette = 0.418), the best among $k \geq 4$.

\begin{table}[htbp]
\centering
\caption{Cluster quality metrics across candidate values of $k$.}
\label{tab:k_selection}
\begin{tabular}{crr}
\toprule
$k$ & \textbf{WCSS} & \textbf{Silhouette} \\
\midrule
2 & 300,965 & 0.5904 \\
3 & 235,336 & 0.4749 \\
4 & 200,088 & 0.3546 \\
\textbf{5} & \textbf{175,597} & \textbf{0.4176} \\
6 & 155,723 & 0.4085 \\
7 & 140,604 & 0.3971 \\
8 & 130,595 & 0.3796 \\
\bottomrule
\end{tabular}
\end{table}

Table~\ref{tab:cluster_profiles} shows the five cluster profiles.

\begin{table}[htbp]
\centering
\caption{Cluster profiles ($k{=}5$). Income and loan amounts are in \$K.}
\label{tab:cluster_profiles}
\begin{tabular}{crrrrrr}
\toprule
\textbf{Cluster} & \textbf{Size} & \textbf{Avg Income} & \textbf{Avg Loan} & \textbf{Avg DTI} & \textbf{Avg CLTV} & \textbf{Denial \%} \\
\midrule
0 & 48,783 & 95 & 206K & 41.7 & 88.8 & 17.0 \\
1 & 20,223 & 225 & 373K & 31.4 & 80.5 & 10.0 \\
2 & 11,314 & 80 & 150K & 63.6 & 70.0 & 79.0 \\
3 & 15,680 & 104 & 116K & 34.5 & 42.7 & 24.0 \\
4 & 3,524 & 519 & 843K & 30.6 & 75.4 & 11.0 \\
\bottomrule
\end{tabular}
\end{table}

Cluster 2 captures debt-overloaded applicants (DTI 63.6\%) with a 79\% denial rate.
Cluster 4 represents affluent applicants (\$519K income) with only 11\% denial.
Clusters 0 and 1 are moderate- and upper-middle-income groups with denial rates of 17\% and 10\%.
Cluster 3 has low CLTV (42.7\%), indicating high-equity borrowers, but a 24\% denial rate.

\begin{figure}[htbp]
\centering
\includegraphics[width=\textwidth]{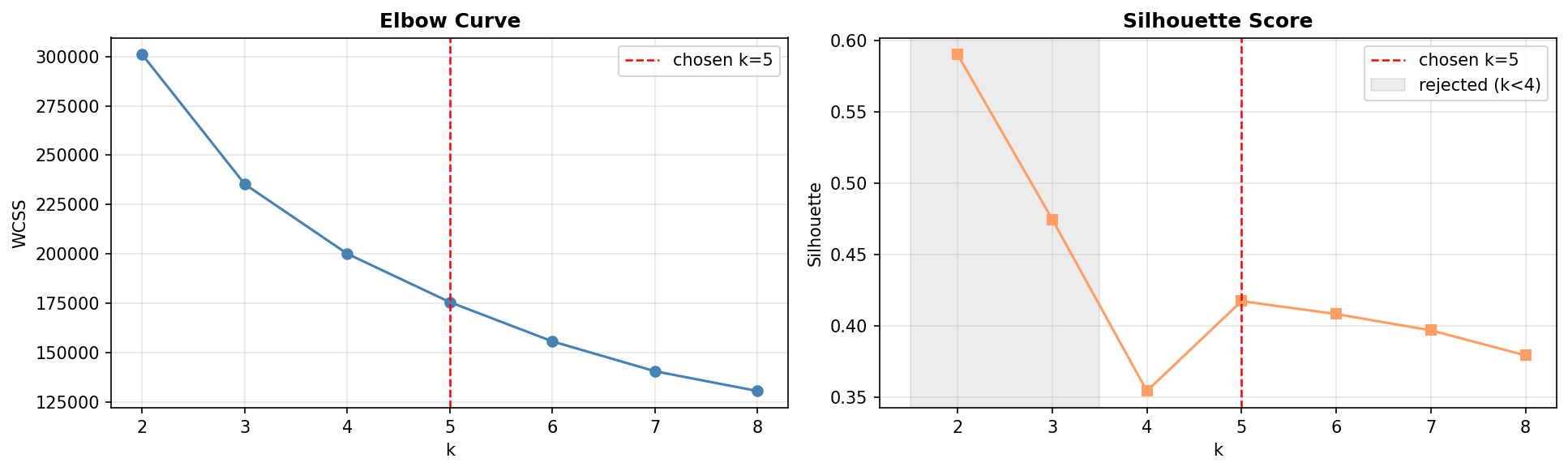}
\caption{Elbow curve (left) and silhouette score (right) for $k$ selection. The gray region shows rejected values below $k{=}4$.}
\label{fig:k_selection}
\end{figure}

\subsection{Disparate Impact Audit}

We audited 45 cluster-group pairs across race, ethnicity, and sex.
Ten pairs were flagged as showing disparate impact (DIR $<$ 0.80), all involving race.
Table~\ref{tab:dir_results} shows the most severe findings.

\begin{table}[h]
\centering
\caption{Top disparate impact findings sorted by DIR (ascending). DIR is computed as the ratio of the protected group's acceptance rate to the reference group's acceptance rate. DIR below 0.80 indicates disparate impact under the four-fifths rule.}
\label{tab:dir_results}
\small
\begin{tabular}{cllrrrrl}
\toprule
\textbf{Cluster} & \textbf{Attribute} & \textbf{Group} & \textbf{Grp Den\%} & \textbf{Ref Den\%} & \textbf{DIR} & \textbf{Grp Size} & \textbf{Flag} \\
\midrule
2 & Race & Native Hawaiian / PI      & 94.12 & 76.81 & 0.254 &    34 & DISPARATE\_IMPACT \\
3 & Race & 2+ minority races         & 51.52 & 20.32 & 0.608 &    33 & DISPARATE\_IMPACT \\
3 & Race & Native Hawaiian / PI      & 51.35 & 20.32 & 0.611 &    37 & DISPARATE\_IMPACT \\
2 & Race & 2+ minority races         & 85.71 & 76.81 & 0.616 &    63 & DISPARATE\_IMPACT \\
2 & Race & American Indian / AN      & 85.16 & 76.81 & 0.640 &   128 & DISPARATE\_IMPACT \\
2 & Race & Black or African American & 84.91 & 76.81 & 0.651 & 2,372 & DISPARATE\_IMPACT \\
2 & Race & Asian                     & 84.84 & 76.81 & 0.654 & 1,075 & DISPARATE\_IMPACT \\
3 & Race & Black or African American & 44.78 & 20.32 & 0.693 & 2,146 & DISPARATE\_IMPACT \\
1 & Race & American Indian / AN      & 29.55 &  9.01 & 0.774 &    44 & DISPARATE\_IMPACT \\
3 & Race & American Indian / AN      & 38.26 & 20.32 & 0.775 &   115 & DISPARATE\_IMPACT \\
\midrule
0 & Race & 2+ minority races         & 30.88 & 14.11 & 0.805 &   136 & ok \\
4 & Race & Black or African American & 26.40 & 10.21 & 0.820 &   125 & ok \\
1 & Race & Black or African American & 24.98 &  9.01 & 0.824 & 1,301 & ok \\
3 & Ethnicity & Hispanic or Latino   & 35.73 & 22.01 & 0.824 & 2,141 & ok \\
2 & Ethnicity & Hispanic or Latino   & 82.26 & 78.67 & 0.832 & 2,435 & ok \\
\bottomrule
\end{tabular}
\end{table}

\begin{table}[h]
\centering
\caption{DIR summary by demographic attribute across all audited cluster-group pairs.}
\label{tab:dir_summary}
\begin{tabular}{lrrrr}
\toprule
\textbf{Attribute} & \textbf{Pairs Audited} & \textbf{Mean DIR} & \textbf{Min DIR} & \textbf{Max DIR} \\
\midrule
Race      & 25 & 0.814 & 0.254 & 1.059 \\
Ethnicity & 10 & 0.932 & 0.824 & 1.006 \\
Sex       & 10 & 1.094 & 1.002 & 1.411 \\
\bottomrule
\end{tabular}
\end{table}

The mean DIR for race-based audits is 0.814.
For ethnicity (Hispanic vs. Not Hispanic), the mean DIR is 0.932.
For sex, the mean DIR is 1.094, indicating no systematic gender-based disparate impact in our data.

In Cluster 3, Black applicants face a 44.8\% denial rate while White applicants face 20.3\% (DIR = 0.693). Both groups fall in the same financial-profile cluster (cluster mean: income \$104K, DTI 34.5, CLTV 42.7), so this gap cannot be explained by the clustering features alone.

\begin{figure}[htbp]
\centering
\includegraphics[width=\textwidth]{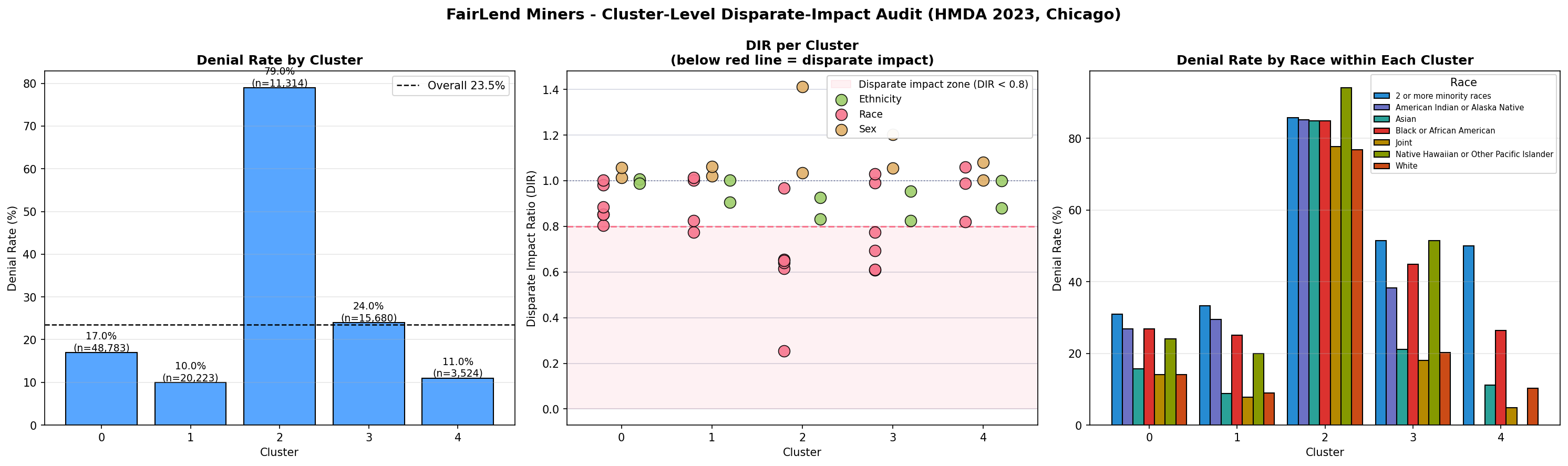}
\caption{Left: denial rate by cluster. Center: DIR values per cluster (below red line = disparate impact). Right: denial rate by race within each cluster.}
\label{fig:dir_audit}
\end{figure}

\section{Discussion}

\subsection{Key Findings}

Our results support six main findings.

First, standard equal-frequency binning introduces 9.63\% racial bias in income discretization.
This matches the 8--10\% reported by Asudeh et al.~\cite{asudeh2025} on credit scoring data, validating the relevance of the unbiased binning problem on HMDA data.

Second, $\varepsilon$-biased binning with seven race groups is infeasible at tight thresholds.
The D\&C algorithm only succeeded at $\varepsilon{=}0.08$ (income) and $\varepsilon{=}0.06$ (loan amount), with high Price of Fairness (29.4\% and 23.5\%).
The paper's experiments primarily use binary groups, and our results confirm that multi-group fairness constraints dramatically reduce the set of feasible solutions.

Third, FP-Growth denial rules are dominated by DTI and are identical under both binning regimes.
At the 10\% support threshold, racial bias does not manifest as explicit high-support patterns.

Fourth, the absence of demographic denial rules does not mean absence of bias.
It means the bias operates indirectly through income and DTI distributions that correlate with race.

Fifth, within-cluster DIR analysis reveals systemic disparate impact.
Ten of 45 audited cluster-group pairs show DIR below 0.80, all involving race.
This discrimination persists even after controlling for income, loan amount, DTI, and CLTV.

Sixth, the two mining approaches are complementary.
FP-Growth identifies the dominant denial factors (DTI) across the whole population, while clustering reveals that racial disparities persist even among financially similar applicants.

\subsection{Limitations}

Several limitations should be noted.

First, HMDA does not include credit score, which is a primary factor in lending decisions; our clustering cannot fully separate discrimination from legitimate risk without this variable.

Second, the 10\% support threshold for FP-Growth is too high to capture minority-group patterns. Lowering it would surface demographic denial rules but also increase spurious results.

Third, our analysis covers Chicago only results may differ in other metropolitan areas with different demographics and housing markets.

Fourth, the fair binning was implemented with all seven race categories, which is more challenging than the binary setting tested in the original paper.

Finally, K-Means assumes spherical clusters of similar size; our clusters range from 3,524 to 48,783 members, which deviates from this assumption. A Gaussian Mixture Model or hierarchical approach might handle this asymmetry better, at the cost of either soft assignments or quadratic memory.




\section{Conclusion and Future Work}

We built a three-stage pipeline to audit discriminatory lending patterns in HMDA 2023 data for the Chicago metropolitan area.
Standard binning introduces measurable racial bias (9.63\%), which fair binning can reduce, though with practical feasibility constraints for multi-group settings.
FP-Growth reveals that high DTI is the dominant denial predictor, while K-Means clustering combined with a DIR audit uncovers racial disparate impact that persists even among financially similar applicant groups.

Future work could extend this analysis in several directions.
Running fair binning with binary race grouping (Black vs. Non-Black) at $\varepsilon{=}0.03$ would provide a cleaner comparison with near-zero Price of Fairness.
Lowering the FP-Growth support threshold to 3--5\% would surface race-specific denial rules.
Computing extended lift per demographic group would directly quantify discrimination strength in the mined rules.
Applying the same pipeline to multiple MSAs would test whether our findings generalize beyond Chicago.

\paragraph{Project Repository.} The project repository is available at \url{https://github.com/Archit1706/FairLend-Miners}.


\end{document}